\documentclass[prl,aps,showpacs,groupedaddress,twocolumn,floatfix]{revtex4}

\usepackage{amsmath,amsfonts,amssymb}
\usepackage{wrapfig}
\usepackage{graphicx}
\usepackage{bbm}
\usepackage{graphics}
\usepackage{subfigure}

\def \beq {\begin{equation}}
\def \eeq {\end{equation}}
\def \ba {\begin{eqnarray}}
\def \ea {\end{eqnarray}}

\newcommand{\ketbrad}[1]{|#1\rangle\!\langle #1|}

\newcommand{\mean}[1]{\langle#1\rangle}

\def\ket#1{\left| #1\right>}
\def\bra#1{\left< #1\right|}

\renewcommand{\section}[1]{{\em #1}:}

\begin{document}

\title{A quantum dot implementation of the quantum NAND algorithm}
\author{J. M. Taylor} \email{jmtaylor@mit.edu} 
\affiliation{Department of Physics, Massachusetts Institute of Technology, 77 Massachusetts Ave, Building 6c-411, Cambridge, MA  02139}
\begin{abstract}
  We propose a physical implementation of the quantum NAND tree
  evaluation algorithm.  Our approach, based on continuous time
  quantum walks, uses the wave interference of a single electron in a
  heirarchical set of tunnel coupled quantum dots.  We find that the
  query complexity of the NAND tree evaluation does not suffer
  strongly from disorder and dephasing, nor is it directly limited by temperature or restricted dimensionality for 2-d structures.  Finally, we suggest a potential application of this algorithm to the efficient determination of high-order correlation functions of complex quantum systems.
\end{abstract}
\date{\today}
\maketitle

Recently a new quantum algorithm was discovered that evaluates a
binary tree of incommeasurate conditions via the logical NAND gate in
a time faster than the provably best classical
algorithm~\cite{hoyer05,farhi07,childs07,childs07b,ambainis07}.  One construction of this
algorithm relies upon a continuous time quantum walk between coupled
bound states, where the ``answer'' to the posed query is determined by
the reflection or transmission of a single-particle wave-packet
through an idealized, complex scattering
device~\cite{farhi98,farhi07}.  Mesoscopic transport in semiconductor
heterostructures have ideal properties for creating a device to
implement a continuous time quantum walk.  Their long phase coherence
times and lengths for electronic transport allows for novel studies of
the electronic properties of small-scale
structures~\cite{imry98,topinka03}.  Modern fabrication techniques
allow for complex structures with robust controls and fast
manipulation~\cite{vanderwiel03,hayashi03,petta04,vidan04}, and the
key features of the continuous time quantum walk approach (a Fano-like
intereference effect with a non-trivial set of coupled bound states)
have been experimentally demonstrated~\cite{johnson04}.

In this paper we propose an approach to implementing NAND trees by
mimicking the logical structure of the Hamiltonian formulation with a
physical arrangement of quantum dots.  Using a recursive Green's function approach inspired by Ref.~\onlinecite{farhi07}, we find the
quantum speed-up of the quantum NAND algorithm---that it takes a time $O(\sqrt{N})$ for $N$ input bits---is maintained if the phase decoherence rate $1/\tau_\phi$~\cite{imry98} and the disorder (detunings and tunnel couplings) be lower than $t/\sqrt{N}$,
where $t$ is the characteristic strength of interdot tunnel coupling
$t$ and $N$ is the number of bits in the input to the NAND tree.  One
key feature of the present work is the use of a large physical space
($O(N)$ quantum dots are used) to allow for efficient coupling to the
complex quantum system (the ``oracle'') to which the query is posed.
We develop appropriate techniques such that temperature and restricted
dimensionality lead to at most a $O(\log N)$ overhead in the evaluation
of the algorithm.  We conclude with a discussion of the ``quantum-ness'' of the algorithm and consider how such a device can be used to measure many-body properties of
complex quantum systems.

\section{Quantum dot implementation}
\begin{figure}
\centering
\includegraphics[width=3.0in]{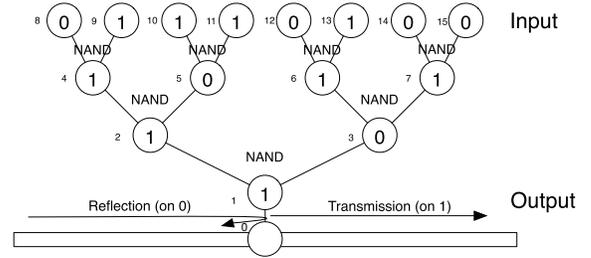}
\caption{
The logical NAND
 tree with $N=8$ inputs in the top of the tree.  Nodes are labeled from
 1 to 15. Each node can be realized as a quantum dot, with the links
 between nodes represented by tunnel coupling between linked dots.
 Detunings $b_i$ applied to dots $i+N$ at the ``top'' of the tree set
the local potential to be off of the Coulomb blockade peak for dot $i+N$ if $b_i \neq 0$.  The conductance of the dot $i=0$ at the bottom of the tree, near zero bias  ($E=0$), encodes the result of the NAND operation.
\label{fig1}}
\end{figure}

The NAND tree quantum walk is described by the transmission
coefficient of a series of coupled states, shown in Fig.~\ref{fig1}.  
Each ``state'' $\ket{i}$ in the graph corresponds to the lowest orbital state of a
quantum dot in the Coulomb blockade regime, with $i=1$ labeling the
bottom quantum dot at level $k=0$ and $i = j+2^k-1$ labeling the $j$th dot at the $k$th level.  We assume a large orbital level spacing $\alpha \gg t$, where $t$ is the average tunnel coupling in the tree of dots.
The coupled dot Hamiltonian can be written:
\begin{equation}
  H = \sum_i \epsilon_i \ketbrad{i} - t_{2i} \ket{i}\bra{2i} -
  t_{2i+1} \ket{i}\bra{2i+1} + {\rm H.c.}
\end{equation}
where $\ket{i} = c^{\dag}_i \ket{n_i}$ describes the addition of a
(spin-less) electron on the $i$th dot near the charge transition from
$n_i \rightarrow n_i +1$.  Each dot can have a slight detuning from
the Fermi energy of $\epsilon_i$, and each dot $i$ is tunnel coupled
to dots $2i$ and $2i+1$ with a tunnel coupling $t_{2i},t_{2i+1} \sim
t$, forming a physical binary tree structure.  

At the top of the tree is an oracle ($\mathcal{O}$), a quantum system with electron
number operators $\hat{n}_{i\mathcal{o}}$ coupled to the binary tree via
$V=\Delta \sum_{i=0}^{N-1} (-)^i \hat{n}_{i\mathcal{O}} \ketbrad{i+N}$.  Unlike previous work, we focus on a coupling that may be easy to implement physically: for
example, the oracle could be a mesoscopic device capacitively coupled
to the top of the tree.  We will proceed with the case of an oracle in an eigenstate of the
$\{\hat{n}_{i\mathcal{O}}\}$ operators with eigenvalues $\{b_i\}$ each taking values 0 or
1.  This sets the detuning of dots at the top of the tree to
$\epsilon_{i+N} = (-)^i b_i \Delta$. 

The algorithm progresses by the injection of an electron wavepacket
into the bottom of the tree.  This probes the Green's function of the
bottom quantum dot, which depends recursively on the Green's functions
for the left and right subtrees.  For wavepackets with small energy
spreads $\Delta E < t/\sqrt{N}$, the successive, coherent scattering
events up and down the tree lead to either total transmission or total
reflection of the wavepacket, where the result (transmission or
reflection) is the final ``answer'' to the NAND query for an input
specified by the bits $b_i$.  A voltage probe in the linear response
regime provides such an energy probe for the system: the conductance
for electrons passing through the bottom of the tree encodes the
result of the tree.  To show that our approach works in the presence
of dephasing and disorder, we develop a recursive description of the
scattering Green's function in the quantum dot tree in the spirit of
previous work~\cite{farhi07}.

\section{Green's function approach}
We now describe an approach to understanding the NAND tree evaluation using a
recursively defined Green's function for single-particle propagation
through the tree.  If we consider a ``tree'' segment with a single
state, $1$, at the bottom, we can define the projectors
$P_1$,$Q_l$,$Q_r$, where $Q_l$ projects into the subspace associated
with the left side of the tree (all states connected to dot 2) and
$Q_r$ similarly projects into the subspace associated with the right side.  We can write the inverse Green's function
exactly~\cite{auerbach}:
\begin{eqnarray}
   G^{-1}_{1}(E) & = & E + i \gamma - P_1HP_1 - \nonumber \\
& & \ \ \sum_{j=l,r}P_1 H Q_j
    \frac{1}{E+i\gamma-Q_jHQ_j} Q_jHP_1 \nonumber \\
 &= & E+i\gamma - \epsilon_{1} - |t_l|^2 G_{l}(E) - |t_r|^2 G_{r}(E) \label{e:recursion}
\end{eqnarray}
This gives a recursive relation for the projected Green's function
$G_{i}(E)$ for a state $i$ in terms of the Green's functions for the
subtrees connected to states $l=2i$ and $r=2i+1$.  The term $\gamma =
1/\tau_\phi$ is a measure of the charge-based dephasing the electron
may be expected to encounter.

At the top (largest fanout) of the tree, we can write the single-state
Green's function
$G^{\rm top}_{i}(E) = \frac{1}{E +i \gamma - \epsilon_i}$
which truncates the recursion relation.  We remark that this recursion
is only successful because the tree contains no loops; otherwise,
Eq.~(\ref{e:recursion}) would fail.   In the disorder-free case,  $t_i = t$
for all $i$ and $\epsilon_i = 0$ except for dots at the top of the tree with input bits $b_k \neq 0$, as described above. 

Setting energy to units of $t$, we consider the Green's function at
the top of the tree for the three different inputs.  We set $s_i$ to
be the sign of the detuning to illuminate the role the alternating
sign $(-)^i$ plays.  We assume $\gamma, E \ll 1 \lesssim \Delta$.  For
dot $j$, tunnel coupled to $2j$ and $2j+1$, we get (to $O(t/\Delta,
\gamma/t)$):
\begin{equation}
G_{j}(E) = \left\{
\begin{array}{ll}
-E/2 - i \gamma / 2  & b_{2j}=0,b_{2j+1}=0 \\
-E - i \gamma   & b_{2j}=0,b_{2j+1}=1\\
\frac{1}{E  + i \gamma + \frac{ s_{2j} + s_{2j+1}}{\Delta} } & b_{2j}=1,b_{2j+1}=1 \end{array} \right.
\label{e:greentree1}
\end{equation}
As the $s_i$ alternate such that each paired ``1'' input has $s_{2j} =
- s_{2j+1}$, the third term has no energy shift.  We now identify
Green's functions of the forms
\begin{equation}
  G_{``0"} = \frac{1}{\alpha E + i \gamma \beta} \ ,\
  G_{``1"} = -(\alpha E + i \gamma \beta) \label{e:logical}
\end{equation}
where the ``0'' and ``1'' indicate the expected NAND results for the
inputs.  Any dependence on $\Delta$ is removed by the alternation of
$s_i$ for paired ``1'' inputs.

We can include effect of disorder in the system, both in variations in
$t_{i}$ and in variations in $\epsilon_i$.  Both
are likely to occur due differences in fabrication, low frequency
noise on electrostatic gates (such as those used to tune the system to
the multi-dot regime) and charge-fluctuators in the system.  For the
three different inputs (00,01,11) at the top of the tree:
\begin{eqnarray}
  0,0 & \rightarrow & - E \frac{ (\epsilon_l - i \gamma)^2 t_l^2 + (\epsilon_r - i
    \gamma)^2 t_r^2}{\left[ \epsilon_r t_l^2 + \epsilon_l t_r^2 - i \gamma (t_l^2 + t_r^2)
    \right]^2} \nonumber \\
  & & \ \ + \frac{1}{\frac{t_l^2}{\epsilon_l - i \gamma} + \frac{t_r^2}{\epsilon_r
        - ig}} \\
  1,0 &\rightarrow & - E \frac{1}{t_r^2} + \frac{\epsilon_r - i \gamma}{t_r^2}
  \\
  1,1 & \rightarrow & \frac{1}{E + (i \gamma + (s_l t_l^2 + s_r
    t_r^2)/\Delta - \epsilon_j)}
\end{eqnarray}
These have the same form as the disorder free results of
Ref.~\onlinecite{farhi07} so long as $t_l, t_r \geq t$, where $t$ is now the
{\rm minimum} tunnel coupling.  More generally, we will assume $t_i$
and $\epsilon_i$ to be drawn from a normal distribution with means
$t,0$ and standard deviations $\sigma_t,\sigma_{\epsilon}$.  We can
remove any explicit $\Delta$ dependence as long as $\sigma_t / \Delta
\ll 1$.

We now show that two subtree's described by Green's functions
of the above type lead to the appropriate type Green's function at the
joining of the two subtrees, recursively implementing the NAND tree.
Consider the general map for two subtrees' Green's functions $G_{l(=2j)}$
and $G_{r(=2j+1)}$, parameterized by $\alpha_{l,r}, \beta_{l,r}$.
$G_{j}$ for the cases 00, 10, and 11 are given to $O(\gamma/t,\epsilon_j/t,t/\Delta)$:
\begin{widetext}
\begin{eqnarray}
  \frac{1}{\alpha_l E + i  \gamma \beta_l},\frac{1}{\alpha_r E
    + i  \gamma \beta_r} &
  \rightarrow &  -\left[\frac{\alpha_l t_r^2 \beta_r^2 + \alpha_r
      t_l^2 \beta_l^2}{(t_r^2 \beta_l + t_l^2 \beta_r)^2} E + i \gamma \frac{\beta_l
      \beta_r}{t_r^2 \beta_l + t_l^2 \beta_r}  \right] \\
  -(\alpha_l E + i  \gamma \beta_l) ,  \frac{1}{\alpha_r E +
    i \gamma \beta_r}&
  \rightarrow &  -\left[ \frac{\alpha_r}{t_r^2} E + i \gamma \frac{\beta_r}{t_r^2} \right]
  \\
  -(\alpha_l E + i \gamma \beta_l) , -(\alpha_r E +
  i \gamma \beta_r) & \rightarrow & \frac{1}{(1+ t_l^2 \alpha_l +
    t_r^2 \alpha_r)E + i \gamma (1 + t_l^2 \beta_l + t_r^2 \beta_r + i
    \epsilon_j / \gamma)}
\end{eqnarray}
\end{widetext}
The overall map implements a NAND gate by identification of ``0'' and
``1'' via Eq.~\ref{e:logical}.  

For a tree consisting entirely of the ``most dangerous case'' (an
input of the form 1011) subtrees of Ref.~\onlinecite{farhi07}, we find
$\alpha \rightarrow 2^{k/2}$, and $\beta \rightarrow 2^{k/2} + i
\sum_{j} \epsilon_j/\gamma$, where the sum is over the set of all dots
where both subtrees evaluate to 1.  The first term indicates that the
Green's function derived is only appropriate within an energy range
$t/2^{n/2} = t/\sqrt{N}$ about $E=0$.  This leads to the condition that the electron wavepacket be spread over a time $\sim \sqrt{N}/t$, and sets $\sqrt{N}$ scaling of the evaluation time.  For random, uncorrelated noise,
the second term (with $O(N)$ terms) has a root-mean-square of
$\sigma_{\epsilon} \sqrt{N}$, and leads to an energy shift by that
amount from the desired resonance at $E=0$ as well as a dephasing
$\gamma 2^{n/2}$.  This indicates that $\gamma, \sigma_{\epsilon} \ll
t/\sqrt{N}$ form the requirements for operation of the device.  Less
difficult is the bounding requirement $t_i \geq t$.  We compared these
results with numerical simulations and find consistent behavior for
tree sizes up to $N=128$.  A sample with $N=32$ is shown in
Fig.~\ref{f:qdconduct}.

\section{Measuring with a quantum dot}
We now detail how a quantum dot coupled to a linear voltage probe can
measure the result of the NAND tree well below the temperature limit
of the leads.  This is a generalization of sub-temperature
spectroscopy used in double quantum dot transport
measurements~\cite{vanderwiel03}.  

We write the Green's function, including the self energies
$\Sigma^R_{l,r}$for left and right leads coupled to the dot via
$t_{l,r}$, as
\begin{equation}
  G_{0}^{-1}(E)  =  E - \epsilon_0 - \Sigma^R_{l} - \Sigma^R_r - 
  |t_1|^2 G_1(E) 
\end{equation}
By working well above the band-bottom of the leads, we can write
$\Gamma = \Gamma_l + \Gamma_r$ with $\Gamma_{l,r} = 2 t_{l,r}^2 \nu =
- 2 i \Sigma^R_{l,r}$, where $\nu$ is the density of states near the
Fermi energy.  The Meir formalism~\cite{meir92} for the transmission
$\bar{T}_{lr}(E) = {\rm Tr}[\Gamma_l G_{l,0} \Gamma_r G_{l,0}^{\dag}]$
yields
\begin{equation}
  \bar{T}_{lr}(E) =  \frac{\Gamma^2/4}{(t_1^2 {\rm Im}[G_1(E)] -
    \Gamma/2)^2 + (E - t_1^2 {\rm Re}[G_1(E)] - \epsilon_0)^2 }
\end{equation}
The conductance follows from the Landauer formula, $G_{lr} = e^2/h
\int f'(E-E_f) \bar{T}_{lr}(E) dE$, where $f(E)$ is the Fermi
distribution.

\begin{figure}
\centering
\includegraphics[width=3.0in]{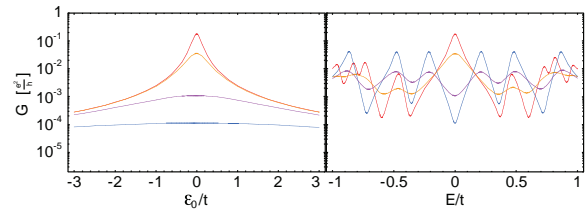}
\caption{
Simulations for $N=32$ with outputs ``0'' (blue,purple) and ``1'' (red,orange).  The dependence of conductance on $\epsilon_0$ (left) and $E$ (right) are shown.  Disorder parameters $\gamma = \sigma_{\epsilon} = \sigma_t = 0.03 t$ (red,blue) and $0.1 t$ (orange,purple) are used. 
\label{f:qdconduct}}
\end{figure}

A few remarks are in order.  First, a pole at $E=0$ for $G_1(E)$ leads
to zero conductance, while a ``1''-like Green's function (proportional
to energy), leads to a peak at $\epsilon_0 = 0$, thus letting the
existence or absence of transport indicate a ``1'' or ``0'' result.
Second, when dephasing ($\gamma$) is weak, the transmission function
is a Lorentzian with a width at most $\Gamma/2$.  When convolved with
the derivative of the Fermi function at $k_b T > \Gamma$, the edges of
the Fermi function are effectively damped, leading to sub-temperature
energy resolution (limited only by $\Gamma$).  To see these effects,
we simulate the expected conductance as a function of $\epsilon_0$ and
$E$ at $T=0$ in Fig.~\ref{f:qdconduct} for an input of $N=32$ bits.
Cases resulting in ``0'' and ``1'' are both shown.  For this scenario,
we set $\Gamma/t = 0.1 < 2^{-5/2}$ and $\Delta/t = 10 > 2^{5/2}$.  The
temperature dependence can be mitigated by further reductions of
$\Gamma$.  However, disorder, as described below, may limit the
achievable resolution.

\section{Two-dimensional layout}
One difficulty encountered for making large NAND trees using this
approach is the increasing perimeter of the tree as a function of tree
depth ($N$ versus $\log N$).  For sufficiently large trees, the
spacing between successive levels $k$ and $k+1$ of the tree must grow
as $2^{k/(d-1)}$ for a $d$-dimensional layout.  Furthermore, tunnel
coupling exponentially decreases as the spacing between quantum dots
increases, potentially limiting this approach to small trees.

\begin{figure}
  \centering
  \includegraphics[width=3.0in]{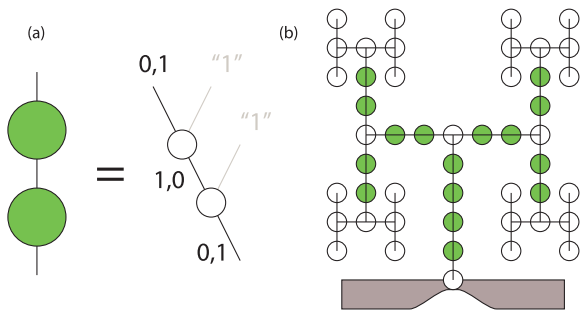}
  \caption{ (a) An inline quantum dot act as an ``inverter''---the
    non-existant right inputs to each dot act as a virtual ``1''.  Two
    inline dots move the bit in the top input to the bottom output.
    (b) An array of tunnel coupled quantum dots in an H-fractal
    layout, with pairs of inverters connecting more distant quantum
    dots.  The number of inverters between dots doubles every {\em
      two} levels of the tree.}
  \label{f:hfractal}
\end{figure}

As an alternative, ``dummy'' levels of the tree can be inserted: these
are states coupled only to a single dot going up or down the tree
(Fig.~\ref{f:hfractal}a).  Each acts as an inverter, where the
``missing'' state is equivalent to the absence of a resonance in one
``leg'' of the tree, which is the ``1''-type Green's function.
Specifically, a state of either ``0'' or ``1'' type parameterized by
$(\alpha,\beta)$ becomes, after coupling through two inverters, $(1 +
\alpha, 1 + \beta)$.  Going a distance $d$ (through $2 d$ inverters)
gives
\begin{equation}
  (d + \alpha, d + \beta)
\end{equation}

As an explicit construction, we consider the H-fractal
(Fig.~\ref{f:hfractal}b) for the layout of quantum dots.  The H-fractal
is a binary tree structure comprised of an ``H'' with the two sides
the same length as the middle bar.  \footnote{A slight modification is
  necessary, as the pair of inverters means all distances (measured in
  dot spacings) must be odd, which in turn requires that each doubling
  of the fractal (moving two levels in the tree) increases the number
  of intervening dots by an even number.  This gives the sequence
  $0,2,4,10,20,38,76, \ldots$, and an area-to-dot size ratio of $3^n$
  for an $n$-level tree.}  This has the benefit that the distance
between successive levels of the binary tree doubles for every {\em
  two} levels.  A ``worst case'' trees (again, with inputs 1011) has
the mapping
\begin{equation}
(\alpha_{k+2},\beta_{k+2}) = (2^{k/2} + 2 \alpha,2^{k/2} + 2 \beta)
\end{equation}
Where the distance goes as $2^{k/2}$ at the $k$th level (with $n$ the
bottom of the tree).  A tree comprised entirely of worst-case trees
has
\begin{eqnarray}
\alpha_n=\beta_n & = & 2^{n/2} + 2 (2^{n/2-1} + 2 (
\ldots (1) \ldots )) \\
& < & n 2^{n/2} = \log_2(N) \sqrt{N}
\end{eqnarray}
This indicates that a 2D architecture leads to an overhead of
$\log(N)$ for the evaluation of the tree.  This still performs better
than the best classical algorithm.  As a final remark, we can
selectively remove one of the pair of dots to produce a ``NOT''
operation inline with the following NAND operation by opening the
tunnel coupling between the pair of dots.  This allows the tree
evalutor to evaluate {\em arbitrary} boolean functions, not just NAND
trees.

\section{Practical limits}
Taking values $\gamma = 0.1 \mu$eV, $t = 100 \mu$eV, $\alpha=1$ meV,
$\Gamma = 0.1 \mu$eV, $\sigma_\epsilon = \sigma_t = 1 \mu$eV, and $k_b
T = 2 \mu$eV (20 mK dilution refrigerator electron temperature), we
find the limiting algorithm size $N \sim 2^{13}$, where the limit is
due to the disorder in the system.  Using the H-fractal layout, this
corresponds to a mesoscopic region of area $a \times 3^{13} \lesssim
0.02\ $mm$^2$ for dot center-to-dot center spacing $a \sim 100$ nm.
This includes appropriate space for electrostatic gates or etched
barriers.  Phase coherent transport over such large regions have been
demonstrated in a number of experiments~\cite{topinka03}.  The
expected evaluation time is of order $10/\Gamma \lesssim 60$ ns (to
generate a differential signal of ten electrons, sufficient for fast
detection).

We can use this NAND tree evaluator as a subsystem in a larger
processing unit.  The best known classical algorithm evaluates in
$N^{0.753}$ time.  Using the quantum device as a sub-tree evaluator of
a quantum oracle, a problem of size $2^{13+k}$ can be evaluated in
$2^{6.5+0.753 k} < 2^{9.79 + 0.753 k}$ time, as subtrees of size $\leq
2^{13}$ can be calculated sequentially using the quantum subsystem as
a hardware accelerator.  Better hybrid quantum-classical approaches
may exist, but require further investigation.

\section{Conclusion}
This paper illustrates a fundamental and heretofore unrecognized
quality of the quantum NAND algorithm.  In particular, the entire NAND
quantum walk can be done with a single-electron wavepacket and
appropriate resonances (quantum dots).  Other
approaches using classical waves, such as light in coupled cavities,
would in principle have the same Green's function.  Thus, from a
quantum information perspective, the interesting part of the algorithm
lies in the oracle: the quantum mechanical system which ``sets'' the
inputs.  We heretofore considered the oracle to be in an eigenstate of
the number operators $\hat{n}_{i\mathcal{O}}$.  If we describe the boolean
function evaluation of the tree by $f(b_0,\ldots,b_{N-1})=0$ or 1, an
arbitrary oracle state leads to an expected result of the tree of
$\mean{f(\hat{n}_{0\mathcal{O}},\ldots,\hat{n}_{N-1\mathcal{O}})}_{\mathcal{O}}$.  For
example, a two-level tree evaluates
$\mean{\hat{n}_{0\mathcal{O}}\hat{n}_{1\mathcal{O}}+\hat{n}_{2\mathcal{O}}\hat{n}_{3\mathcal{O}}-\hat{n}_{0\mathcal{O}}\hat{n}_{1\mathcal{O}}\hat{n}_{2\mathcal{O}}\hat{n}_{3\mathcal{O}}}$, a non-trival, high-order correlation function of the oracle.  We now have the
outstanding task of identifying interesting quantum systems to use as an oracle.

\begin{acknowledgments}
  The author thanks E. Farhi, J. Goldstone, S. Gutmann, C. Marcus, and S. Jordan, and the workshop on Solid State Quantum Information Systems at the Niels Bohr International Academy.  This research is supported by the Pappalardo Fellowship.
\end{acknowledgments}

\end{document}